\documentclass[11pt,cite]{article}
\usepackage{latexsym}

\usepackage[latin1]{inputenc}

\usepackage{graphicx}
\usepackage{color,pst-plot}

\usepackage{amsmath}
\usepackage {amsfonts,amssymb,amstext}

\newcommand{\lbl}[1]{\label{eq:#1}}
\newcommand{ \rf}[1]{(\ref{eq:#1})}

\newcommand{\be}{\begin{equation}}
\newcommand{\ee}{\end{equation}}
\newcommand{\bea}{\begin{eqnarray}}
\newcommand{\eea}{\end{eqnarray}}
\newcommand{\setl}{\setlength\arraycolsep{2pt}}

\newcommand{\noi}{\noindent}
\newcommand{\nn}{\nonumber}
\newcommand{\ra}{\rightarrow}

\newcommand{\cO}{{\cal O}}

\newcommand{\Imm}{\mbox{\rm Im}}

\newcommand{\annd}{\mbox{\rm and}}

\newcommand{\gL}{\frac{1-\gamma_{5}}{2}}
\newcommand{\gR}{\frac{1+\gamma_{5}}{2}}

\newcommand{\stern}{\langle\bar{\psi}\psi\rangle}

\input epsf

\def\theequation{\arabic{section}.\arabic{equation}}

\begin{document}

\begin{titlepage}

\begin{flushright}
\end{flushright}

\vspace*{0.2cm}
\begin{center}
{\Large {\bf On Anomaly Matching and Holography}}\\[2 cm]

{
{\bf Marc Knecht}~$^a$, {\bf Santiago Peris}~$^b$ and {\bf Eduardo de Rafael}~$^a$}\\[1cm]

$^a$  {\it Centre  de Physique Th{\'e}orique~\footnote{Unit{\'e} Mixte de Recherche (UMR 6207) du CNRS et des Universit{\'e}s Aix Marseille 1, Aix Marseille 2 et Sud Toulon-Var, affili{\'e}e {\`a} la FRUMAM.}\\
       CNRS-Luminy, Case 907\\
    F-13288 Marseille Cedex 9, France}

\vspace*{1.0cm}

$^b$ {\it Grup de F{\'\i}sica Te{\`o}rica\\ Universitat Aut\`onoma de Barcelona\\ 08193 Barcelona, Spain} \\[0.5cm]

\end{center}

\vspace*{3.0cm}

\begin{abstract}
We discuss the possible validity in QCD of a relation between Green's functions which has been recently suggested by Son and Yamamoto, based on a class of AdS/CFT-inspired models of QCD. Our conclusion is that the relation in question is unlikely to be implemented in QCD.
\end{abstract}

\end{titlepage}

\section{\normalsize Introduction}\lbl{int}
\setcounter{equation}{0}
\def\theequation{\arabic{section}.\arabic{equation}}

\noi
Since the work by Maldacena~\cite{Mal99} there has appeared a wealth of papers in the literature based on holographic AdS/QCD-inspired models with various  phenomenological claims.  In Refs.~\cite{SonStephanov1, SonStephanov2}, an attempt was made to identify the Kaluza-Klein states of a five dimensional quantum  field theory with the infinity of meson resonances obtained in QCD in the large-$N_c$ limit. Although tantalizing, this identification is not without shortcomings. For instance, no asymptotically-free beta function was obtained and, although in some particular cases the parton model logarithm was reconstructed for the short distance behavior of two-point functions, the full condensate expansion at large momentum was missing. Furthermore, a linear spacing rule was obtained for the meson masses rather than their squares, unlike the case of the Regge trajectories. It was then claimed~\cite{Erlich} that one could obtain this infinite spectrum for large-$N_c$ QCD from a Pad\'{e} approximation to the parton model logarithm, an approach that had been proposed many years earlier in Ref.~\cite{Migdal}. This is certainly a rather courageous statement as there are an infinity of spectra which are consistent with the same parton model logarithm at high momentum~\cite{Cata}. A non trivial dilaton field was later on introduced in Ref.~\cite{Karch} as a possible mechanism to recover the right Regge behavior in the spectrum.

 The lack of a proper Operator Product Expansion  (OPE) was tackled in Ref.~\cite{Hirn}. In this reference, it was recognized that, in order to accomplish matching with the OPE in QCD, it is necessary to assume a different five-dimensional metric for the vector and for the axial channels, giving up on the existence of a single gravity dual, which was the original motivation of the approach. Moreover, it has been recently pointed out in Ref.~\cite{Ambrosio} that a consistent treatment of the vector and tensor channels precludes the introduction of dilatonic backgrounds as a mechanism to achieve the correct QCD-like Regge behavior of the spectrum, calling into question the viability of the aforementioned mechanism proposed in \cite{Karch}. Other properties of QCD, such as its typical jet-like structure in parton collisions, also show disagreement with those based on theories with a  gravity dual~\cite{Csaki}.

 Given this state of affairs, we think it would be extremely interesting if one could show the claimed gravity equivalence in the one case in which the solution of large-$N_c$ QCD is known, namely in two dimensions. Although in ~\cite{Katz} some progress has been made along these lines, the full solution was not found. This could also help establish the connection with the light-front formulation discussed in~\cite{Brodsky}.

We think it is fair to say that, up until now, the concrete proposals show a very large model dependence, which seriously questions whether one is truly learning about real QCD.  It is crucial that some predictions may be found for which there is some type of universality that could put them on a firmer basis, independently of the particular gravity model chosen. Fortunately, one such prediction has recently appeared in the literature~\cite{SY10}.

The authors of \cite{SY10} claim that, in  a whole class of theories whose gravity dual is described by the Yang-Mills-Chern-Simons theory with chiral symmetry broken by boundary conditions in the infrared, the following relation can be considered as a \emph{generic} result (see Eqs.~\rf{pi} and \rf{triangle} below):
\be\lbl{SY}
 w_L (Q^2) -2 w_T (Q^2) = -\frac{2N_c}{f_{\pi}^2} \Pi_{\rm LR}(Q^2)\, ,
\ee
and that this result holds, at least approximately, in real QCD. Moreover, in the other class of theories with a scalar field representing the chiral condensate~\cite{SonStephanov2} the previous result does not follow and, according to~\cite{SY10}, this second class of theories should be ruled out as QCD-like candidates, at least in their simplest setting. Since this  second class are sort of complementary to the first, this leaves us with the validity of the result \rf{SY} as a clear-cut test of whether, as a matter of principle, the present AdS/CFT-related ideas can be useful for QCD or not. Consequently, in this note we would like to discuss the validity of ~\rf{SY} for QCD.

Since the relation \rf{SY} is a highly non--trivial one, let us first review what is known in QCD about the various Green's functions which appear in it.

In the r.h.s. $\Pi_{\rm LR}(Q^2)$  denotes the self-energy of the familiar correlation function
\be\lbl{PiLR}
\Pi_{\rm LR}^{\mu\nu}(q)=2i\int d^4 x\,e^{iq\cdot x}\langle 0\mid
T\left(L^{\mu}(x)R^{\nu}(0)^{\dagger}\right)\mid 0\rangle \,,
\ee
of  left and right currents:
\be
L^{\mu}(x)=\bar{d}(x)\gamma^{\mu}\frac{1}{2}(1-\gamma_{5})u(x)
\qquad \annd \qquad
R^{\mu}(x)=\bar{d}(x)\gamma^{\mu}\frac{1}{2}(1+\gamma_{5})u(x)\,.
\ee
In the chiral limit where the light quark masses are set to zero ($Q^2=-q^{2}\ge 0$
for
$q^2$ spacelike)
\be\lbl{pi}
\Pi_{\rm LR}^{\mu\nu}(q)=(q^{\mu}q^{\nu}-g^{\mu\nu}q^2)\Pi_{\rm LR}(Q^2)\,,
\ee
and the self--energy function
$\Pi_{\rm LR}(Q^2)$ vanishes order by order in perturbation theory; it becomes  an
order parameter of the spontaneous breakdown of chiral symmetry for all
values of the momentum transfer~\cite{FSS90,KdeR98}. Unless otherwise stated we shall be working in the chiral limit.

The low $Q^2$ behaviour of $\Pi_{\rm LR}(Q^2)$ is
governed by the effective chiral Lagrangian of QCD, i.e. the Lagrangian
formulated in terms of the Goldstone degrees of freedom and external local
sources only and, in the large-$N_c$ limit, reads:
\be
-Q^2\Pi_{\rm LR}(Q^2)\underset{{Q^2\ \ra 0}}{\thicksim}\ f_{\pi}^2+4L_{10} Q^2+8 C_{87} Q^4+\cO(Q^6)\,,
\ee
where $f_{\pi}$ is the pion coupling constant (the same $f_{\pi}$ which appears in the r.h.s. of Eq.~\rf{SY})  and
$L_{10}$ and $C_{87}$ denote specific coupling constants of $\cO(p^4)$ and $\cO(p^6)$ of the
low energy effective chiral Lagrangian~\cite{GL85,ABT00}. For later use, let us note, at this stage, that the right-hand side of Eq.~\rf{SY} can also be expressed, in terms of the function ${\widehat \Pi}_{\rm LR}(Q^2)$ defined by
\be\lbl{Pihat}
{\widehat \Pi}_{\rm LR}(Q^2) \,=\, -\,\frac{1}{3} \, g_{\mu\nu} \Pi_{\rm LR}^{\mu\nu}(q) \,=\, -Q^2\Pi_{\rm LR}(Q^2) ,
\ee
as
\be\lbl{SYbis}
-\frac{2N_c}{f_{\pi}^2} \Pi_{\rm LR}(Q^2) \,=\, + \lim_{m\rightarrow 0}\frac{2N_c}{Q^2}\, \frac{{\widehat\Pi}_{\rm LR}(Q^2)}{{\widehat\Pi}_{\rm LR}(0)} ,
\ee
for all $Q^2$. This way of writing Eq.~\rf{SY} also stresses the fact that it is not a linear relation between Green's functions.

The high $Q^2$ behaviour of $\Pi_{\rm LR}(Q^2)$  is governed by the operator product expansion~\cite{SVZ79} and, in the large-${ N_c}$ limit, one obtains
\be\lbl{OPE}
\Pi_{\rm LR}(Q^2)\underset{{Q^2\ \ra \infty}}{\thicksim}\ -4\pi^2 \left(\frac{\alpha_s}{\pi}+\cO(\alpha_s^2)\right)\stern^2 \frac{1}{Q^6}+ \cO\left(\frac{1}{Q^8}\right)\,.
\ee

On the other hand,
the functions $ w_L (Q^2)$ and $w_T (Q^2)$ which appear in the l.h.s. of Eq.~\rf{SY} are the longitudinal and transverse functions of the VVA triangle of electroweak hadronic currents in a specific kinematic configuration~\cite{KPPdeR04}:
\be\lbl{triangle}
Q^2 \left[w_L (Q^2) -2 w_T (Q^2) \right]=\frac{16\pi^2}{\sqrt{3}}\int d^4 x \int d^4 y e^{iq\cdot x}(x-y)_{\lambda}
\epsilon^{\mu\nu\rho\lambda}\langle 0\vert \hat{T}\left\{L_{\mu}^3 (x) V_{\nu}^3 (y) R_{\rho}^8 (0) \right\}\vert 0\rangle\,,
\ee
where ($\lambda_i$ are flavour $SU(3)$ Gell-Mann matrices)
\be
L_{\mu}^3 (x)=\bar{\psi}(x)\frac{\lambda_3 }{2}\gamma_{\mu}\gL\psi(x)\,, \quad R_{\rho}^8 (0)=\bar{\psi}(0)\frac{\lambda_8 }{2}\gamma_{\rho}\gR\psi(0)\,, \quad V_{\nu}^3(y)=\bar{\psi}(y)\frac{\lambda_3 }{2}\gamma_{\nu}\psi(y)\,,
\ee
and $\hat{T}$ denotes the appropriate prescription for the chronological product~\cite{KPPdeR04}.
The combination of functions $w_L (Q^2) -2 w_T (Q^2)$ in the chiral limit is also an order parameter of spontaneous chiral symmetry breaking. The same combination of functions appears naturally in the calculation of the contribution to the muon anomaly from the VVA triangle of electroweak  currents~\cite{KPPdeR02,CMV03,KPPdeR04}.

The Adler-Bell-Jackiw anomaly fixes $w_L (Q^2)$ at all values of $Q^2$:
\be\lbl{anomaly}
w_L (Q^2)=2\frac{N_{c}}{Q^2}\,,
\ee
with $N_c$ the number of colors (the same $N_c$ as in Eq.~\rf{SY}).

As for the transverse function $w_T (Q^2)$ the present situation in QCD is as follows: at the one loop level in perturbation theory (pQCD)
\be\lbl{vain}
w_T ^{\rm\footnotesize pQCD}(Q^2)=\frac{N_c}{Q^2}
\ee
and, surprisingly,  as first shown by Vainshtein~\cite{Vain03} and subsequently confirmed in ref.~\cite{KPPdeR04},
this result remains valid to all orders in pQCD. However, as first shown in ref.~\cite{KPPdeR02},  $w_T (Q^2)$ receives non--perturbative QCD contributions and the result in Eq.~\rf{vain} ceases to be valid as one enters moderate and low $Q^2$ values. In fact, at large-$Q^2$ values and in the  large-${N_c}$ limit~\cite{KPPdeR02,CMV03}
\be\lbl{SDWT}
w_T (Q^2)\underset{{Q^2\ \ra \infty}}{\thicksim}\ \frac{N_c}{Q^2}- 32\pi^4 \left(\frac{\alpha_s}{\pi}+\cO(\alpha_s^2)\right)\stern\ \Pi_{\rm VT}(0) \frac{1}{Q^6}+\cO\left(\frac{1}{Q^8} \right)\,,
\ee
where $\Pi_{\rm VT}(0)$ denotes the invariant function of the {\bf v}ector--{\bf t}ensor correlation function at zero momentum transfer:
\be
\int d^{4}y e^{ik\cdot y}\langle 0\vert T\left\{\bar{\psi}\gamma_{\sigma}\frac{\lambda^{a}}{2}
\psi(y)\ \bar{\psi}\sigma^{\beta\delta}\frac{\lambda^{b}}{2}\psi(0)\right\}\vert 0\rangle=(k^{\beta}\delta_{\sigma}^{\delta}-k^{\delta}
\delta_{\sigma}^{\beta})\delta^{ab}\Pi_{\rm VT}(k^2)\,.
\ee
At small $Q^2$ values~\cite{KPPdeR02}
\be
w_T (Q^2)   \underset{{Q^2\ \ra 0}}{\thicksim}\ 128\pi^2 C_{22}^{W}+\cO(Q^2)\,,
\ee
with $C_{22}^{W}$ one of the $\cO(p^6)$ low--energy constants of the effective chiral Lagrangian in the odd--parity sector~\cite{BGT02}.

The SY--relation in Eq.~\rf{SY} is claimed to be valid at all $Q^2$ values. We will now discuss its consequences for low and high-$Q^2$.

\section{\normalsize The SY--Relation at Short and Long Distances.}
\setcounter{equation}{0}
\def\theequation{\arabic{section}.\arabic{equation}}

\noi
We observe that at large-$Q^2$ values the leading $\frac{1}{Q^2}$ behaviours of $w_L (Q^2)$ and $w_T (Q^2)$ in the combination  $w_L (Q^2)-2w_T (Q^2)$  cancel out and we are left, both in the l.h.s. and in the r.h.s. of Eq.~\rf{SY},  with leading terms which are $\cO\left(\frac{1}{Q^6} \right)$. Unfortunately, the calculation of the residue~\cite{KPPdeR02} of the $\cO\left(\frac{1}{Q^6} \right)$ term in  $w_T (Q^2)$ involves the unknown parameter $\Pi_{\rm VT}(0)$ which, so far, can only be estimated with models. The SY-relation in Eq.~\rf{SY}, with neglect of higher order $\alpha_s$ corrections, would imply
\be\lbl{SYVT}
\Pi_{\rm VT}(0)\vert_{\rm SY}= \frac{1}{8\pi^2}\frac{N_c}{f_{\pi}^2}\stern\ .
\ee
Notice, however, that the result for the OPE in Eq. ~\rf{SDWT} has not really been obtained with the five-dimensional gravity theory in \cite{SY10}. In particular, no tensor fields were considered in this reference, which leaves the function $\Pi_{\rm VT}$ appearing in \rf{SDWT} out of reach. The gravity dual theory only relates the two sides of the Eq. ~\rf{SY} for any $Q^2$. Were one really to compute the OPE from the dual theory, one would obtain at best an exponential fall-off in $Q^2$ beyond the parton model logarithm, as explicitly demonstrated in the Appendix of \cite{SY10}.  In fact, in Ref. \cite{Ambrosio}, an analysis of the combined set of sum rules for the vector and tensor channels in the context of holographic models has been undertaken. The conclusion of this analysis is that both the mechanism for linear confinement suggested in \cite{Karch} and the usual AdS/CFT prescription are incompatible with these sum rules. As we emphasized in the introduction, the problem of reproducing the condensate expansion remains.

Concerning the long--distance behaviour of the SY--relation in Eq.~\rf{SY}, the leading $\frac{1}{Q^2}$ term from $w_{\rm L}(Q^2)$ in the l.h.s.  exactly cancels the leading term on the r.h.s.  and, to first non--trivial order, one is left with the equality
\be\lbl{LESY}
C_{22}^{W}\vert_{\rm SY}= - \frac{N_c}{32 \pi^2 f_{\pi}^2}L_{10}\,.
\ee
Unfortunately, contrary to the coupling $L_{10}$ which is well known phenomenologically, there is no model independent determination of the constant $C_{22}^{W}$ . We notice, however, that  Eq.~\rf{LESY}
is a rather strange one since it relates a coupling of the parity odd sector in the effective chiral Lagrangian to another
coupling which is in the parity even sector. We therefore suggest to investigate the issue under discussion from yet another
point of view.

\vspace*{1cm}

\section{\normalsize The SY--Relation in perturbative QCD (pQCD).}
\setcounter{equation}{0}
\def\theequation{\arabic{section}.\arabic{equation}}

Equation \rf{SY} implies a relationship between Green's functions. If valid in QCD it should also be formally valid in pQCD. Of course, we know that the evaluation of Green's functions using pQCD at long distances ceases, in general,  to reproduce hadronic physics correctly;  but an equation among Green's functions should also hold when using pQCD to evaluate the two sides of that equation. This is precisely what we want to examine next. Although this is a rather academic exercise, we think it is nevertheless a valid one if the goal is an assessment of the validity of identities such as \rf{SY}.

Using \rf{anomaly} and \rf{vain}, and given that  \rf{PiLR} identically vanishes in pQCD in the chiral limit, one might initially think that \rf{SY} is trivially satisfied. However, $f_{\pi}$ in the denominator also vanishes  in pQCD in this limit, so the validity of \rf{SY} is, in fact, far from obvious. One should first discuss how this chiral limit is taken, and this requires the consideration of a nonvanishing quark mass.

Given a non-zero quark mass $m$, and some convenient UV regulator\footnote{We will take dimensional regularization, with $D=4-\epsilon$, for simplicity.}, evaluation of the Green's function \rf{triangle} in pQCD leads to the result (for $m^2\ll Q^2$) \cite{Melnikov}
\be\lbl{Melni}
\lim_{\epsilon\rightarrow 0} \Big[w_L(Q^2;m,\epsilon)- 2\ w_T(Q^2;m,\epsilon)\Big]=\frac{N_c}{Q^2}\ \left(\frac{N_c^2-1}{2 N_c}\right)\frac{\alpha_s}{\pi}\  \frac{m^2}{Q^2}\ \left(2 \log\frac{Q^2}{m^2}+ 1\right)+ \mathcal{O}(m^4/Q^4,\alpha_s^2 )\ .
\ee
One then has that
\be\lbl{limit}
\lim_{m\rightarrow 0}  \lim_{\epsilon\rightarrow 0} \Big[w_L(Q^2;m,\epsilon)- 2 w_T(Q^2;m,\epsilon)\Big]=0 \ ,
\ee
 as expected  since, as already stated, the combination
 $w_L(Q^2)- 2 w_T(Q^2)$ is an order parameter of spontaneous chiral symmetry breaking~\cite{KPPdeR04} .

In full QCD, the correlation function $\Pi_{\rm LR}^{\mu\nu}(q)$ in the presence of light quark masses ($m_u =m_d =m$ for simplicity), depends in general on two invariant functions~\cite{FNdeR79}:
\be
\Pi_{\rm LR}^{\mu\nu}(q)=(q^{\mu}q^{\nu}-g^{\mu\nu}q^2)\Pi_{\rm LR}^{(1)}(Q^2)+q^{\mu}q^{\nu}\Pi_{\rm LR}^{(0)}(Q^2)\, .
\ee
In the chiral limit one has that $\Pi_{\rm LR}^{(1)}=\Pi_{\rm LR}$ in Eq. \rf{pi}, and $\Pi_{\rm LR}^{(0)}=0$

The result in Eq.~\rf{SY} means, on account of the relation \rf{SYbis}, that  the double limit in \rf{limit} can also be computed on the following combination of vacuum polarization functions, in the presence of a quark mass $m$ and UV regulator $\epsilon$,\footnote{We do not explicitly write the dependence on $m$ and $\epsilon$ in $\Pi_{\rm LR}^{(0,1)}$ for simplicity of notation.}
\be\lbl{SYp}
\lim_{m\rightarrow 0}  \lim_{\epsilon\rightarrow 0} \left\{\frac{2 N_c}{Q^2}  \ \frac{{\widehat\Pi}_{\rm LR}(Q^2)}{ {\widehat\Pi}_{\rm LR}(0)}\right\}\ .
\ee
This result should also vanish to agree with \rf{limit}. Equivalently, the equality

$$ w_L (Q^2) -2 w_T (Q^2)=\left\{\frac{2 N_c}{Q^2}  \ \frac{{\widehat\Pi}_{\rm LR}(Q^2)}{ {\widehat\Pi}_{\rm LR}(0)}\right\}$$
proposed in ref. \cite{SY10}, has to be verified order by order in a simultaneous expansion in $\alpha_s$ and $\epsilon$ in the regulated bare theory, as any Ward identity is supposed to do, but only to $\mathcal{O}(m^0)$ since its validity is limited to the chiral limit. However,  using lowest order perturbation theory, we will now explicitly see that this is not the case.

An elementary calculation of the one-loop diagram yields $\Pi_{\rm LR}^{(1)}(Q^2)=-\Pi_{\rm LR}^{(0)}(Q^2)$, with the following  result
{\setl
\bea\lbl{explicit}
\Pi_{\rm LR}^{\mu\nu}(q) & = & 2i\ N_c\ \nu^{-\epsilon}\int \frac{d^D p}{(2\pi)^D}\frac{2m^2 g^{\mu\nu}}{(p^2-m^2+i\epsilon)[(p-q)^2 -m^2+i\epsilon]}\
\nn \\
 & = & -\frac{N_c}{4\pi^2}\left\{\frac{2}{\epsilon}-\gamma_{\rm E} +\log 4\pi
 -\int_0^1 dx \log\left[\frac{m^2+x(1-x)Q^2-i\epsilon}{\nu^2}\right] \right\} m^2 g_{\mu\nu}\nn \\
 & = & -\frac{N_c}{4\pi^2}m^2 g_{\mu\nu}\ \left\{\frac{2}{\epsilon}-\gamma_{\rm E} +\log 4\pi -\log\frac{m^2}{\nu^2}\right. \nn \\
  & & \left.  +2+\sqrt{1+\frac{4m^2}{Q^2}}\log
  \left[\frac{\sqrt{1+\frac{4m^2}{Q^2}}-1}{\sqrt{1+\frac{4m^2}{Q^2}}+1} \right]\right\}\,.
\eea}

\noi
This same result may also be obtained from an unsubtracted dispersion relation  with the following spectral functions\footnote{It is interesting to notice the presence of the delta-function term to account for the UV divergence.}
{\setl
\bea\lbl{spect}
\frac{1}{\pi}\Imm\Pi_{\rm LR}^{(1)}(t)&=&-\frac{1}{\pi}\Imm\Pi_{\rm LR}^{(0)}(t)=\\
&&\!\!\!\!\!\!\frac{N_c}{4\pi^2}
\frac{m^2}{t}\sqrt{1-\frac{4m^2}{t}}\ \theta(t-4m^2) -\frac{N_c}{4\pi^2}m^2\ \left\{\frac{2}{\epsilon}-\gamma_{\rm E} +\log 4\pi -\log\frac{m^2}{\nu^2}\right\}\delta(t)\,. \nn
\eea}

Let us now use the above expression \rf{explicit} to compute the combination appearing in \rf{SYp}. One obtains (for $m^2\ll Q^2$)
{\setl
\bea\lbl{result}
&& 2 \ \frac{N_c}{Q^2} \lim_{m\rightarrow 0} \lim_{\epsilon \rightarrow 0}  \left\{\frac{{\widehat\Pi}_{\rm LR}(Q^2)}{ {\widehat\Pi}_{\rm LR}(0)}\right\}= \nn\\
&& =\ 2\  \frac{N_c}{Q^2} \lim_{m\rightarrow 0}
\lim_{\epsilon\rightarrow 0}\frac{\frac{2}{\epsilon}-\gamma_{\rm E} +\log 4\pi  -\log\frac{m^2}{\nu^2}+ 2 +\log\frac{m^2}{Q^2}+ \mathcal{O}\left( \frac{m^2}{Q^2}\log\frac{m^2}{Q^2} \right) }{\frac{2}{\epsilon}-\gamma_{\rm E} +\log 4\pi  -\log\frac{m^2}{\nu^2}}\nn \\
&&=\ 2\ \frac{N_c}{Q^2}
\eea}

\noindent
which does \emph{not} vanish and is leading at large $N_c$. In principle, one could expect that quark mass corrections, which are chirally suppressed nonperturbatively, could fix this problem\footnote{We thank D. T. Son for discussions on this point.}. However, this would require two wonders rather than one, as the mismatch between \rf{Melni} and \rf{result} involves not only the powers of the quark mass but also the powers of $\alpha_s$. Whether this is possible or not would
require full knowledge of the equation \rf{SY} away from the chiral limit.We conclude that Eq.~\rf{SY} is unlikely to be an identity in QCD, even in the large-$N_c$ limit.

\vspace*{1cm}

\section{\normalsize Conclusion.}
\setcounter{equation}{0}
\def\theequation{\arabic{section}.\arabic{equation}}

On the basis of the previous analyses we conclude that the SY-relation is unlikely to hold in QCD, at least for the wide class of models considered in \cite{SY10}, even when the large-$N_c$ limit is taken. Whether \rf{SY} may still be considered valid in some ``approximate''  dynamical  sense will completely depend on the type (and size) of the corrections one expects to \rf{SY}. Without this knowledge, the usefulness of \rf{SY} is, regretfully, very limited. However, we fail to  imagine what this dynamical approximation may possibly be. Our conclusion, therefore, is that, unless a major breakthrough takes place, the so-called ``AdS/QCD'' approach is very unlikely to teach us about properties of QCD.

\vspace*{3cm}

\begin{center}
{\normalsize\bf Acknowledgements.}
\end{center}

We would like to thank D.T. Son and N. Yamamoto for discussions. One of us (SP) would like to thank M. Golterman for comments on the manuscript and O. Cat\`{a} for extensive discussions about AdS/QCD. This work has been partially supported by  CICYT-FEDER-FPA2008-01430, CPAN CSD2007-00042 and SGR2009-00894.

\vspace*{0.5cm}

\begin{center}
{\normalsize\bf Note added in proof.}
\end{center}

While this paper was considered for publication, there have appeared two other papers related to this subject \cite{Colangelo:2011xk,Iatrakis:2011ht}.


\vfill

\end{document}